# *Free-space* creation of *ultralong anti-diffracting light beam* with *multiple energy oscillations* adjusted using *'optical pen'*

Xiaoyu Weng[1]*, Qiang Song[3], Xiaoming Li[1], Xiumin Gao[2], Hanming Guo[2]*, Junle Qu[1]* and Songlin Zhuang[2]

*1 Key Laboratory of Optoelectronic Devices and Systems of Ministry of Education and Guangdong Province, College of Optoelectronic Engineering, Shenzhen University, Shenzhen, 518060, China.*

*2 Engineering Research Center of Optical Instrument and System, Ministry of Education, Shanghai Key Lab of Modern Optical System, School of Optical-Electrical and Computer Engineering, University of Shanghai for Science and Technology, 516 Jungong Road, Shanghai 200093, China.*

*3 Optics Department, Telecom Bretagne, Institut Mines-Telecom, CS83818, 29238 Brest Cedex 03, France.*

* e-mail address: *xiaoyu@szu.edu.cn*; *hmguo@usst.edu.cn*; *jlqu@szu.edu.cn*.

# Abstract:

A light beam propagating with an infinite anti-diffracting distance requires infinite power to preserve its shape. However, the fundamental barrier of finite power in free space has made the problem of diffraction insurmountable over the past few decades. Here, to overcome this limitation, we report an approach that employs the multiple energy oscillation mechanism, thereby permitting the creation of a light beam with an ultralong anti-diffracting distance in free space. The anti-diffracting distance is no longer restricted by finite power in free space but instead depends on the number of energy oscillations. This unprecedented propagation behavior is attributed to a new understanding of non-diffractive light beam: when an anti-diffracting light beam completely discharges its energy, it cannot recharge again. A versatile 'optical pen' is therefore developed to manipulate the number, amplitude, position and phase of energy oscillations for an arbitrary numerical aperture of a focusing lens so that energy recharge can occur in free space and multiple energy oscillations can be realized. A light beam with a tunable number of energy oscillations is eventually generated in free space and propagates along a wavy trajectory. This work will enable extending non-diffractive light beams to an expanded realm and facilitate extensive developments in optics and other research fields, such as electronics and acoustics.

# Introduction

Diffraction is a natural property of light beam, which tends to make it broaden during propagating in free space. Fighting against diffraction effect is one of the most important topics in the community of optics. However, not until 1987, the advent of Bessel beam brings the hope of addressing successfully this issue [1]. Great success has been achieved in the study of anti-diffracting light beams since then [1-13]. Every breakthrough regarding these light beams always leads to the development of many applications, including optical imaging [14-16], optical trapping [17-19], optical communication [20-22] and laser-assisted guiding of electric discharges [23]. All advances are associated with the long anti-diffracting distances of these light beams. For many years, the scientific consensus has been that ideal diffraction-free light beams require infinite power to maintain their shapes during propagation in free space [6, 7, 11]. Nevertheless, the finite power in free space makes the occurrence of light beams with ultralong anti-diffracting (UAD) distances impossible [See Supplementary Information: Section (1)]. Although great efforts have been made to generate light beams with long non-diffractive distances, such as the optical needle [11-13], Bessel beam [1, 2], and Airy beam [6-9], all share very limited anti-diffracting distances in free space. Among them, Airy beams possess the longest anti-diffracting distances, though they are no longer than $32\lambda$ in free space for a numerical aperture (*NA*) of 0.8; see Fig. S13.

The optical nonlinearity induced by the interaction between light beams and special materials can be used as an alternative to suppress diffraction effect so that light beams can preserve their shapes without divergence over a long distance [24-29]. For example, light beams that propagate over more than 1000 Rayleigh lengths can be achieved due to the nonlinear response of dipolar glass [29]. A spatial light soliton can also propagate invariantly in other nonlinear materials, where the diffraction effect can be compensated by the self-focusing effect [24-28]. Nevertheless, since such light beams can exist only inside nonlinear material in which optical nonlinearity occurs, their wide utilization in many applications is limited. For example, light beams in nonlinear material cannot excite the fluorescent material inside a cell during the imaging process [14, 15]. Although air has been demonstrated as a valid nonlinear material to support light bullets, their anti-diffracting distances are still restricted by finite aperture, namely finite power in free space [30, 31]. Undoubtedly, the generation of light beams with UAD distances in free space, not just in a particular material, is of great value but still far beyond our understanding.

Here, we report an approach that overcomes this fundamental barrier by employing a multiple energy oscillation mechanism to generate light beams with UAD distances in free space. The non-diffractive distance of a UAD light beam is determined only by the number of energy oscillations, rather than finite power in free space. The transition from power-dependent to power-independent anti-diffracting distance is attributed to a new understanding of anti-diffracting light beam based on energy oscillation mechanism: when an anti-diffracting light beam completely discharges its energy, it cannot recharge again. A versatile 'optical pen' is therefore developed to manipulate the number, amplitude, position and phase of energy oscillation so that energy recharge can occur in free space and multiple energy oscillations can be realized. Eventually, UAD light beams with a tunable number of energy oscillations can be obtained in free space.

# Results

An ideal anti-diffracting light beam can propagate without divergence limitlessly in free space; thus, infinite power is necessary to preserve its shape. However, being restricted by the finite power in free space, until now, only a light beam with a finite non-diffractive distance has been obtained in practice [6-9]. It is debatable whether a light beam with a UAD distance can be created in free space. The key to this problem is the new physical insight regarding the propagation of anti-diffracting light beam in free space.

**Energy Oscillation: a general property of anti-diffracting light beam**

In theory, anti-diffracting light beams can be considered as the Fourier transformation of their corresponding Fourier spectrums [1, 4, 6, 7]. For example, cubic phase and high-pass pupil filter are the Fourier spectrums of Airy beam [6, 7] and Bessel beam [1], respectively. The process of Fourier transformation can be flexibly achieved by an objective lens [32]. Thus, all anti-diffracting light beams can be created in the focusing system of Fig. S1. According to the general focusing theory in Eq. (3s), one can readily obtain the energy fluxes of anti-diffracting light beams at two symmetrical points $P_1(x, y, z)$ and $P_2(x, y, -z)$, which can be expressed as [see Eq. (8s)]

$$\langle \mathbf{S} \rangle_{tP_1} = -\langle \mathbf{S} \rangle_{tP_2} \tag{1}$$

Thus, when propagating in free space, the light beam experiences two inverse energy processes that transfer the energy from $\langle \mathbf{S} \rangle_{tP_2}$ to $\langle \mathbf{S} \rangle_{tP_1}$. Here, $\langle \mathbf{S} \rangle_{tP_2}$ and $\langle \mathbf{S} \rangle_{tP_1}$ are called as energy charge and discharge, respectively.

Energy charge and discharge compose an entire energy oscillation, which is a directional energy flux that confines the energy into an interaction between mainlobe and sidelobes. Thus, the light beams would not diverge freely like that of Gaussian beam in free space. For example, a quasi-Airy beam exhibits one energy oscillation from mainlobe to sidelobes and then from sidelobes to mainlobe in Fig. S4, while Bessel beam inverses in Fig. S7. If only energy oscillation occurs, all anti-diffracting light beams can preserve their shapes without divergence. Even when encountering an obstacle, the mainlobe can carry out self-healing with the power from sidelobes [8]. This is the reason why anti-diffracting light beams are naturally composed with mainlobe and sidelobes. Moreover, since all those light beams can be created by focusing their corresponding Fourier spectrums in Fig. S1, energy oscillation is therefore a general property for all anti-diffracting light beams.

**Conceptual Change via Energy Oscillation Mechanism**

Based on energy oscillation in Eq. (1), an anti-diffracting light beam can be divided into two symmetrical parts according to whether the energy flux z<0 or z>0. Both parts experience two different energy processes: energy charge (z<0) or energy discharge (z>0) that prevent the light beam from diverging in free space. In principle, the non-diffractive distance can be magnified by increasing the strength of energy oscillation. However, as already demonstrated in Fig. S5 and S6, the energy charge process cannot be strengthened limitlessly because of the finite power in free space. A finite energy charge can support only a finite energy discharge, thus leading to the impossibility of creating a light beam with UAD distance for one particular energy oscillation in free space. Although this power barrier in free space is insurmountable, the energy oscillation mechanism still offers a new possibility to transform this power-dependent distance into a power-independent one. Again, an anti-diffracting light beam experiences a finite energy charge when z<0 as well as a finite energy discharge when z>0. When the light beam completely discharges its energy (z>0), it cannot be recharged again. Thus, the diffraction effect finally dominates, and the light beam can no longer propagate. For this reason, the solution to this problem is no longer restricted by the finite power in free space but depends on the ability to recharge after the energy is completely discharged.

**Critical condition for Energy Recharge**

Although energy oscillation is a general property shared by all anti-diffracting light beams, it doesn't mean that all those light beams are suitable to recharge energy again when finishing energy discharge in free space. Energy recharge occurs only at the switch point between adjacent energy oscillations, where one energy oscillation is almost finished performing energy discharge while another is just beginning to recharge. Thus, the energy flux in the former energy oscillation must be the same as that of the latter one at the switch point. Otherwise, the incontinuity of the energy flux will cause interference between adjacent energy oscillations.

Supposed there are two modes of light beams, the energy fluxes of which can be expressed as

$$\begin{aligned}\left\langle \mathbf{S}_{\delta 1}\right\rangle_{tP_2} &= -\left\langle \mathbf{S}_{\delta 1}\right\rangle_{tP_1} \\ \left\langle \mathbf{S}_{\delta 2}\right\rangle_{tP_2} &= -\left\langle \mathbf{S}_{\delta 2}\right\rangle_{tP_1}\end{aligned} \quad (2)$$

When the first mode ($\delta 1$) of light beam almost finishes energy discharge $\left\langle \mathbf{S}_{\delta 1}\right\rangle_{tP_1}$, the second mode ($\delta 2$) must recharge the energy again so that the light beam can remain anti-diffracting. As a result,

$$\langle \mathbf{S}_{\delta 1} \rangle_{tP_1} = \langle \mathbf{S}_{\delta 2} \rangle_{tP_2} \tag{3}$$

From Eq. (2, 3), one can readily obtain the critical condition for energy recharge in free space, which can be simplified as

$$\langle \mathbf{S}_{\delta 1} \rangle_{tP_2} = -\langle \mathbf{S}_{\delta 1} \rangle_{tP_1} = -\langle \mathbf{S}_{\delta 2} \rangle_{tP_2} = \langle \mathbf{S}_{\delta 2} \rangle_{tP_1} \tag{4}$$

That is, only light beams satisfied Eq. (4) can be utilized to recharge energy again in free space. As already demonstrated in Supplementary Information: Section (2), this critical condition can be easily satisfied by quasi-Airy beams with $\pm\eta$, which can be expressed as [see Eq. (11s)]

$$\langle \mathbf{S}_{\eta} \rangle_{tP_2} = -\langle \mathbf{S}_{\eta} \rangle_{tP_1} = -\langle \mathbf{S}_{-\eta} \rangle_{tP_2} = \langle \mathbf{S}_{-\eta} \rangle_{tP_1} \tag{5}$$

Thus, a quasi-Airy beam with $\eta$ can be recharged by its mutually complementary mode $-\eta$.

**UAD Light Beam with Multiple Energy Oscillations**

Here, we introduce the multiple energy oscillation mechanism to recharge the energy of a quasi-Airy beam after energy discharge, thus permitting the generation of UAD light beams in free space. This energy mechanism, just as its name implies, requires multiple energy oscillations in the focal region of the lens, as shown in Fig. 1. For the $i$-th energy oscillation, the cubic phase plate can be expressed as [see Eq. (9s)]

$$T_{ci} = \exp\left\{i\eta_i \frac{k \sin^{\sigma_i} \theta}{\sin^{\sigma_i} \alpha} \left[\sin^3(\varphi + \phi_0) + \cos^3(\varphi + \phi_0)\right]\right\} \tag{6}$$

where $\eta_i$ denotes the period of the cubic phase plate and $\sigma_i$ is the parameter that controls the distribution of the cubic phase plate; in this paper, $\sigma_i=3$. $T_{ci}$ stands for the standard cubic phase. The whole phase can be rotated by an angle $\phi_0$. $\theta$ and $\varphi$ are the convergence angle and azimuthal angle, respectively. $\alpha=arcsin(NA/n)$, where $NA$ is the numerical aperture of the lens and $n$ is the refractive index in free space.

Since energy recharge occurs only at the switch point between adjacent energy oscillations, where one energy oscillation is almost finished performing energy discharge while another is just beginning to recharge, the parabola-like energy oscillations, which descend when $\eta > 0$ and ascend when $-\eta$, must come in pairs so that the energy of $\langle \mathbf{S}_{\eta} \rangle_{tP_2}$ can be recharged by $\langle \mathbf{S}_{-\eta} \rangle_{tP_1}$, while $\langle \mathbf{S}_{-\eta} \rangle_{tP_2}$ can be recharged by $\langle \mathbf{S}_{\eta} \rangle_{tP_1}$. Accordingly, energy recharge can be realized by simply overlapping $\langle \mathbf{S}_{\eta} \rangle_{tP_2}$, $\langle \mathbf{S}_{-\eta} \rangle_{tP_2}$ with their corresponding $\langle \mathbf{S}_{-\eta} \rangle_{tP_1}$, $\langle \mathbf{S}_{\eta} \rangle_{tP_1}$ at the switch point.

However, this process leads to a highly precise manipulation of each energy oscillation in the focal region of the lens. An 'optical pen' is therefore developed to manipulate the position, number, amplitude and phase of each energy oscillation simultaneously and precisely so that energy recharge can occur at the switch point and multiple energy oscillations can be realized in free space, the phase of which can be written as [see Eq. (19s)]

$$\psi_p = Phase\left[\sum_{i=1}^{N} PF(s_i, x_i, y_i, z_i, \delta_i)\right] \quad (7)$$

where $N$ indicates the number of foci. $x_i$, $y_i$, and $z_i$ denote the position of the $i$-th focus in the focal region. $s_i$ and $\delta_i$ are the weight factors, which can be used to adjust the amplitude and phase of the $i$-th focus, respectively.

By combining the 'optical pen' in Eq. (7) with the cubic phase plate in Eq. (6), the final phase for the generation of a UAD light beam with multiple energy oscillations can be simplified to

$$\psi_{SND} = Phase\left\{\sum_{i=1}^{N}[T_{ci} PF(s_i, x_i, y_i, z_i, \delta_i)]\right\} \quad (8)$$

where the 'optical pen' is responsible for controlling the number, position, amplitude and phase of the energy oscillations, while the cubic phase determines the orientation of the energy oscillations. This phase can be considered as the Fourier spectrum of UAD light beam, which can easily be implemented using the phase-only spatial light modulator (SLM) shown in Fig. 1.

**Experiment**

Following the general focusing theory of linearly polarized Gaussian beams presented in Eq. (3s), a UAD light beam with multiple energy oscillations can be obtained by substituting $T_c$ for $T_{SND} = \exp(i\psi_{SND})$. Note that energy oscillations exist naturally in free space; thus, a UAD light beam can always be realized with arbitrary $NA$. Without loss of generality, we take $NA=0.095$ as examples to generate UAD light beams with different numbers of energy oscillations in air, namely, $n=1$. Their corresponding results with $NA=0.8$ can be found in Fig. S13. In the following simulations and experiments, the unit of length in all figures is the wavelength $\lambda$, and the light intensity is normalized to the unit value. The anti-diffracting distance of UAD light beam is evaluated with FWHM (full width at half maximum).

Figure 1 presents the schematic of the experimental setup. A collimated incident x linearly polarized Gaussian beam (wavelength: 632.8 *nm*) propagating along the optical axis passes through a phase-only SLM and two lenses (L$_1$, L$_2$) before it is focused by the objective lens (*OL*), with $NA=0.095$. L$_1$ ($f_1=120$ mm) and L$_2$ ($f_2=150$ mm) compose a 4*f*-system that makes the phase coded in the SLM and the entrance plane of the *OL* conjugate. The light intensity of the UAD light beam in the x-y plane is recorded using a CCD, which can be moved along the optical axis by a motorized precision translation stage (z stage). By recording the light intensities in different z planes using the CCD, UAD light beams with different numbers of energy oscillations can be reconstructed. The blue and yellow arrows represent the energy charge and discharge processes, respectively. One pair of adjacent blue and yellow arrows indicates an entire energy oscillation.

Figure 2(i-l) and (e-h) present the experimental and theoretical results of UAD light beams with 0, 1, 2, and 3 energy oscillations, respectively. The phases coded in the SLM are shown in Fig. 2(a-d); the parameters can be found in Table 1. The anti-diffracting distances of the UAD light beams with 0, 1, 2, and 3 energy oscillations in the experiment [Fig. 2(i-l)] are approximately 204.1$\lambda$, 2544.9$\lambda$, 5606.2$\lambda$ and 8316.1$\lambda$, while their corresponding theoretical results in Fig. 2(e-h) are 195.7$\lambda$, 2560$\lambda$, 5592$\lambda$ and 8310$\lambda$, respectively. The deviation ratio between both results is defined as |ADD$_{ex}$-ADD$_{th}$|/ADD$_{th}$, where ADD$_{ex}$ and ADD$_{th}$ are the anti-diffracting distance for the experimental and theoretical UAD light beams, respectively. The maximum deviation ratio of the UAD light beams in Fig. 2 is 4.29%, and the experimental results match well with the theoretical predictions. Generally, a light beam with a longer anti-diffracting distance than that of one energy oscillation has long been considered impossible due to the finite power in free space. However, using the multiple energy oscillation mechanism, UAD light beams can be realized in an energy charge-discharge-recharge-discharge way, the anti-diffracting distance of which can be extensively promoted by increasing the number of energy oscillations via the 'optical pen'.

To better understand the multiple energy oscillation mechanism, we take the UAD light beam shown in Fig. 2(g) as an example and investigate the light intensities and transverse energy fluxes (green arrows) in the $z = -1100\lambda, 0, 1100\lambda$ planes, which are denoted by points A, B, and C, respectively. Their corresponding light intensities are shown in Fig. 3 (A-C), while the energy fluxes are indicated by the green arrows in Fig. 3 (PA-PC), respectively. As shown in Fig. 3(PA-PC), Point A experiences an energy discharge in the initial energy oscillation. Point B is the switch point, at which the initial energy is nearly fully discharged, and the second energy charge is just beginning. Point C falls totally within the second energy charge process. Supposed power compensation does not occur at point B, then the light beam may endure significant divergence and cannot propagate further in free space, similar to the common quasi-Airy beam in Fig. 2(f).

However, the UAD light beam is able to recharge again at point B via the multiple energy oscillation mechanism. Therefore, the sidelobes at point B must be cross-shaped so that the bottom and upper energy fluxes are in the same direction, as shown in Fig. 3(PB). Here, the bottom sidelobe corresponds to an energy discharge process similar to that at point A in Fig. 3(PA) in the initial energy oscillation, while the upper one corresponds to the subsequent energy charge similar to that at point C in Fig. 3(PC) in the second energy oscillation. During propagating in free space, the initial energy oscillation becomes small as the light beam passes through point B and eventually vanishes. As shown in Fig. 3(PC), the second energy oscillation finally takes over, and the light beam undergoes a second energy charge at point C. In this way, a UAD light beam can propagate over a super-long range by only repeating energy oscillations without being restricted by the finite power in free space. Their experimental corresponding light intensities are shown in Fig. 3(a-c). The same energy process also occurs for the UAD light beam with 3 energy oscillations. Points d and f in Fig. 2(l) are two switch points, at which the energy discharge process is almost finished and a new energy charge process is just beginning. Point e in Fig. 3(e) continues to be anti-diffracting in the second energy oscillation process, as predicted in Fig. 3(E). Compared with the theoretical results in Fig. 3(D, F), the cross-shaped sidelobes in Fig. 3(d, f) are slightly asymmetric, which is attributed to the deviation of the light path in the experiment.

**Discussion**

Creating a UAD light beam in free space involves four key points: revealing an underlying mechanism of anti-diffracting light beam, renewing our understanding on the propagation behavior of anti-diffracting light beam, finding a suitable anti-diffracting light beam for energy recharge and developing a "optical pen" for the realization of UAD light beam with multiple energy oscillations. For the first two points, diffraction-free is a general property shared by all anti-diffracting light beams, which can only be explained by their mathematical forms derived from Helmholtz Equation in the past. However, by exploring the energy flux of those light beams, energy oscillation is demonstrated to be the physical reason of this common property, which confines the energy of anti-diffracting light beams into an interaction between mainlobe and sidelobes so that they won't diverge freely like that of Gaussian beam during propagating in free space. This mechanism not only renews our understanding on the propagation behavior of anti-diffracting light beams, but also provides a possibility to fight against diffraction effect in free space. That is, when anti-diffracting light beams finish energy discharge, they cannot be recharged energy again. This conceptual change paves the way for creating UAD light beams with the multiple energy oscillation mechanism in free space.

For the third point, despite the general energy oscillation mechanism, it does not mean that all those light beams are suitable to be recharged energy again in free space. One critical condition must be satisfied, that is, the energy fluxes of the initial energy discharge must be equal to those of the secondary energy charge at the switch point between adjacent energy oscillations. Thus, only light beams satisfied Eq. (4) can be recharged in free space. For example, the Bessel beam at point $P_1$ cannot be recharged by the energy flux of point $P_2$ using the 'optical pen' due to their inverse energy fluxes [see Supplementary Information Figure S7]. Overlapping the energy flux of point $P_1$ with that of point $P_2$ will cause an incontinuity of energy fluxes at the switch point, thereby leading to interference at the switch point. Thus, it is impossible to create UAD light beams with the Bessel beam as their base. In contrast, a quasi-Airy beam possesses a pair of mutually complementary modes, with $\pm\eta$. Eq. (5) shows that a quasi-Airy beam of $\eta$ can be recharged simply by using that of $-\eta$. Hence, a UAD light beam can propagate much further than a Bessel beam [1, 2] and a quasi-Airy beam [6-9] simply by repeating energy oscillations in free space.

For the fourth point, 'optical pen' is another crucial aspect responsible for the realization of the multiple energy oscillation mechanism. As a versatile optical tool, the 'optical pen' has an explicit form (Eq. (7)), which can be used to unify the relationship between the focal pattern and the phase in the entrance plane. Taking the focus array in Fig. S10 as an example, the 'optical pen' represents all possible phases using only two different weight factors $s_i$ and $\delta_i$. Thus, it is flexible to adjust the position, number, amplitude and phase of each focus in the focal region simultaneously and precisely, which ensures that energy recharge can occur at the switch points between adjacent energy oscillations. This advantage makes the 'optical pen' the perfect optical tool for the creation of UAD light beams.

When finishing the above processes, UAD light beam can be built up by concatenating a sequence of energy oscillations with alternate signs of their curvatures. However, it should be emphasized that UAD light beam is a new kind of light beam in free space, not just multiple foci in the focal region. Firstly, as a light beam, the energy flux must evolve continually so that the information won't be lost during propagation. Since Eq. (4) is satisfied, a quasi-Airy beam can be recharged energy at the switch point with its mutually complementary mode. Therefore, the energy flux of UAD light beams can evolve without mutation during propagating in free space. Secondly, UAD light beam can be considered as a sum of Airy beams series with different weight factors, which can be simplified as

$$UAD = \sum_{i=1}^{N} s_i AiB_i(x-\Delta x_i, y-\Delta y_i, z-\Delta z_i) \tag{9}$$

where $AiB_i(x-\Delta x_i, y-\Delta y_i, z-\Delta z_i)$ stands for the *i*-th energy oscillation with the position $(\Delta x_i, \Delta y_i, \Delta z_i)$ and the weight factor $s_i = 1$, which can be adjusted by the optical pen flexibly. As seen from Eq. (9), UAD light beam is a special solution of Helmholtz Equation. Thirdly, UAD light beam propagates in a wavy trajectory, while Airy beam can only propagate parabolically. That is, UAD light beam is not a kind of Airy beam. Based on the above reasons, one can conclude that UAD light beam is an entirely new anti-diffracting light beam in free space.

To some extent, UAD light beam has something in common with the light beams generating in the nonlinear material, such as spatial light solitons. Both kinds of light beams can propagate without diverging by suppressing diffraction effect. However, the way to fight against diffraction effect is totally different. Without taking advantage of light-material interaction, the creation of UAD light beam in free space conveys a physical idea that diffraction effect can be suppressed by the property of anti-diffracting light beam itself, namely energy oscillation. Energy oscillation is a directional energy flux shared by all anti-diffracting light beams, which confines the energy into an interaction between mainlobe and sidelobes. Thus, if only energy oscillations occur, those light beams can preserve their shapes without divergence during propagating in free space. To this point, UAD light beam can be considered as a similar 'spatial optical soliton' of free space.

In conclusion, we have theoretically and experimentally demonstrated that light beams with UAD distances can be generated via the multiple energy oscillation mechanism in free space. The non-diffractive distance is no longer restricted by the finite power in free space but depends on the number of energy oscillations, which can be flexibly manipulated using the 'optical pen'. For example, UAD light beams with 3 energy oscillations can propagate without significant divergence over $8316.1\lambda$ when *NA*=0.095, while they can do so over only $2544.9\lambda$ when *NA*=0.095 for a quasi-Airy beam. Due to their super-long non-diffractive distances, UAD light beams offer promising applications in many scientific studies, such as those on optical imaging [14-16], laser-assisted guiding of electric discharges [23], optical trapping [17-19], and optical communication [20-22].

In addition to the interests in practical applications, this work may have great impact on the community of optics, even other research areas, such as electron and acoustic. Firstly, although the energy oscillation mechanism satisfying Eq. (4) was firstly found for a quasi-Airy beam, other accelerating light beams, such as Mathieu and Weber beam [4], may also possess a similar mechanism. Thus, other kinds of UAD light beams with the multiple energy oscillation mechanism can also be created in free space, which may spark a new era regarding the study of similar "spatial light solitons" in free space. Secondly, since electron Airy beam and self-bending acoustic beam have already been demonstrated in free space [33, 34], one may get inspiration from this work to create their counterpart UAD light beams and solve the problems in both fields.

**Acknowledgments**

Parts of this work were supported by the National Natural Science Foundation of China (61525503/61620106016); National Basic Research Program of China (2015CB352005); Guangdong Natural Science Foundation Innovation Team (2014A030312008); Hong Kong, Macao and Taiwan cooperation innovation platform & major projects of international cooperation in Colleges and Universities in Guangdong Province (2015KGJHZ002); Shenzhen Basic Research Project (JCYJ20150930104948169/JCYJ20160328144746940/GJHZ20160226202139185); and National Key Research and Development Program of China (2016YFF0101603).

**Author contributions**

X. Weng conceived the research and designed the experiments. Q. Song and X. Weng performed the experiments and analyzed all the data. X. Li, X. Gou and H. Guo supervised the experiments. X. Weng wrote the paper, and H. Guo offered advice regarding its development. J. Qu and S. Zhuang directed the whole project.

**Competing interests statement**

The authors declare that they have no competing financial interests.

# Figure

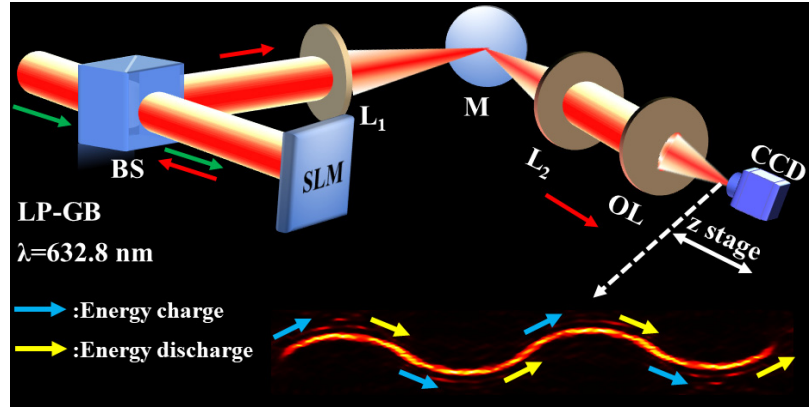

**Figure 1** Schematic of the experimental setup. A collimated incident x linearly polarized Gaussian beam (LP-GB), with a wavelength of $\lambda=632.8$ *nm*, propagating along the optical axis passes through a phase-only SLM and two lenses ($L_1$, $L_2$) before it is focused by the objective lens (*OL*) (*NA*=0.095). $L_1$ ($f_1$=120 mm) and $L_2$ ($f_2$=150 mm) compose a 4*f*-system that makes the phase coded in the SLM and the entrance plane of the *OL* conjugate. UAD light beams with different numbers of energy oscillations can be reconstructed by recording the light intensity in different z planes using a CCD, which can be moved along the optical axis by a motorized precision translation stage (z stage). For example, a UAD light beam with four energy oscillations propagates in an energy charge-discharge-recharge-discharge way, with the blue and yellow arrows denoting energy charge and discharge, respectively. One pair of adjacent blue and yellow arrows denotes an entire energy oscillation. The green and red arrows indicate the propagation directions of incident and modulated beams, respectively.

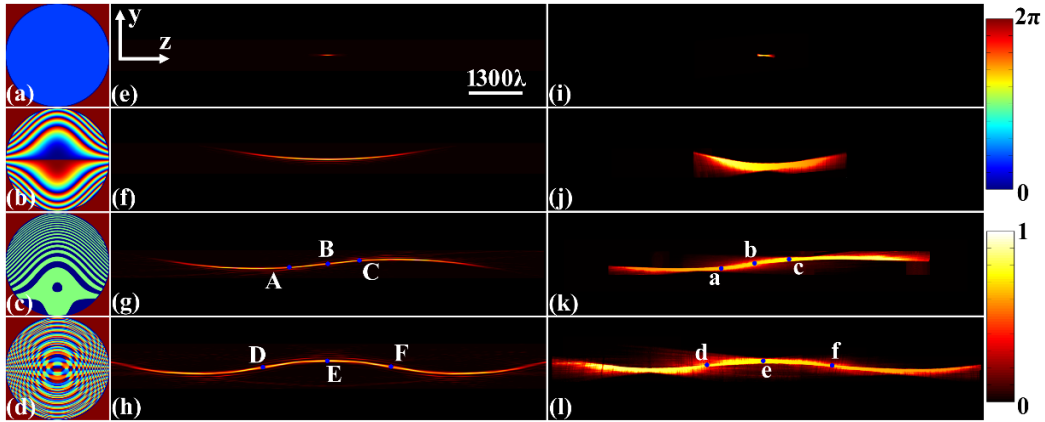

**Figure 2** Theoretical (e-h) and experimental (i-l) results for UAD light beams with 0, 1, 2, and 3 energy oscillations under the condition of *NA*=0.095, which are generated by the phases in (a-d), respectively. The anti-diffracting distances in the experiment are approximately 204.1$\lambda$, 2544.9$\lambda$, 5606.2$\lambda$ and 8316.1$\lambda$ (i-l), and their corresponding theoretical results (e-h) are 195.7$\lambda$, 2560$\lambda$, 5592$\lambda$ and 8310$\lambda$, respectively. All parameters of the phases can be found in Table 1.

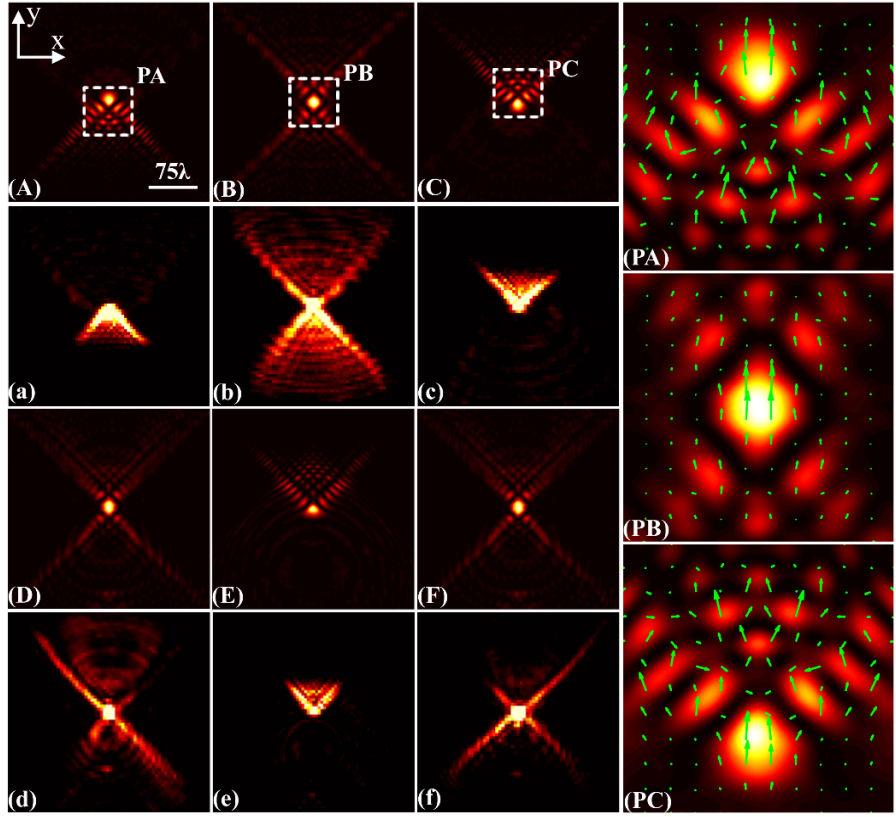

**Figure 3** Theoretical (A-F) and experimental (a-f) results for UAD light beams in different z planes, where (a, A) $z=-1100\lambda$; (b, B) $z=0$; (c, C) $z=1100\lambda$; (d, D) $z=-1465\lambda$ (e, E) $z=0$, and (f, F) $z=1465\lambda$. All are denoted by the points a-f and A-F in Fig. 2, respectively. Points b, d, f and B, D, F are the theoretical and experimental switch points, respectively. (PA-PC) are the energy fluxes of (A-C), respectively, which are indicated by the green arrows.

| Table 1: UAD Light Beams of Figure 2 | | | | | | |
|---|---|---|---|---|---|---|
| $N$ | $y_1$ | $y_2$ | $y_3$ | $z_1$ | $z_2$ | $z_3$ |
| 1 | 0 | | | 0 | | |
| 1 | 0 | | | 0 | | |
| 2 | -42.5 | 42.5 | | -1535 | 1535 | |
| 3 | -42.5 | 42 | -42.5 | -3046 | 0 | 3046 |
| $\eta_1$ | $\eta_2$ | $\eta_3$ | $\delta_1$ | $\delta_2$ | $\delta_3$ | Parameter |
| 0 | | | | | | $NA = 0.095$ |
| -5 | | | | | | $\phi_0 = 0.75\pi$ |
| -5 | 5 | | $-0.07\pi$ | $0.07\pi$ | | $x_i = 0$ |
| -5 | 5 | -5 | $0.1\pi$ | 0 | $-0.1\pi$ | $s_i = 1$ |

# Supplementary Information for

# Free-space creation of ultralong anti-diffracting light beam with multiple energy oscillations adjusted using 'optical pen'

Xiaoyu Weng[1]*, Qiang Song[3], Xiaoming Li[1], Xiumin Gao[2], Hanming Guo[2]*, Junle Qu[1]* and Songlin Zhuang[2]

1 *Key Laboratory of Optoelectronic Devices and Systems of Ministry of Education and Guangdong Province, College of Optoelectronic Engineering, Shenzhen University, Shenzhen, 518060, China.*

2 *Engineering Research Center of Optical Instrument and System, Ministry of Education, Shanghai Key Lab of Modern Optical System, School of Optical-Electrical and Computer Engineering, University of Shanghai for Science and Technology, 516 Jungong Road, Shanghai 200093, China.*

3 *Optics Department, Telecom Bretagne, Institut Mines-Telecom, CS83818, 29238 Brest Cedex 03, France.*

* e-mail address: *xiaoyu@szu.edu.cn*; *hmguo@usst.edu.cn*; *jlqu@szu.edu.cn*.

# 1. The definition of "Power" for anti-diffracting light beam

In principle, an anti-diffracting light beam can be generated by focusing its Fourier spectrum [1-4], namely incident light beam with corresponding wavefront modulation. Thus, it is easy to confuse the "power" of incident light beam before the objective lens in Fig. S1 with that of anti-diffracting light beam mentioned in the paper. For the former one, the power of incident light beam has not impact on the anti-diffracting distance, but the brightness of anti-diffracting light beam in the focal region. Although longer anti-diffracting distance gives rise to weaker brightness of anti-diffracting light beam, the normalized light intensity remains the same. Thus, the light intensity of UAD light beam is normalized to the unit value.

For the latter one, the power mentioned in the paper is the energy of anti-diffracting light beam itself, which can be obtained by integrating the amplitude along the propagating trajectory. As shown in Fig. S2(a), an ideal Bessel beam with amplitude proportional to $J_n(k_r r)\exp(-ik_z z)$ can preserve its shape without divergence infinitely during propagating in free space. Thus, the power of ideal Bessel beam can be obtained by [1]

$$I_B = \left| \iint_{-\infty}^{+\infty} J_n(k_r r)\exp(-ik_z z) dr dz \right|^2 \to \infty \tag{1s}$$

Likewise, the power of ideal Airy beam in Fig. S2(b) can be obtained by [2]

$$I_A = \left| \iint_{-\infty}^{+\infty} Ai(s - \xi^2/4)\exp[i(s\xi/2 - \xi^3/12)] dx dz \right|^2 \to \infty \tag{2s}$$

where $s = x/x_0$ and $\xi = z/kx_0^2$, $x_0$ is an arbitrary constant. Since Bessel and Airy function are not square integrable, infinite power is therefore needed to maintain theirs shapes. However, this can only be achieved by the infinite aperture of the lens. In practice, finite aperture can only transfer finite power to anti-diffracting light beam, thereby leading to a finite anti-diffracting distance in free space. From this point of view, creation of UAD light beam are always considered to be a hopeless task in free space.

## 2. Energy oscillation: a general property of anti-diffracting light beam

Anti-diffracting light beams are special solutions of Helmholtz Equation (HE). According to their mathematical forms, those light beams can preserve their shapes without divergence infinitely during propagating in free space [1-4]. However, due to the finite power in free space, only light beams with finite anti-diffracting distance can be obtained. It is well known that anti-diffracting light beams are composed of mainlobe and sidelobes. As the light intensity of sidelobes increases, the anti-diffracting distance increases accordingly. Up to now, little is known about the mechanism hide behind this physical phenomenon. To reveal this mechanism, one must know what role do the sidelobes and mainlobe play in fighting against the diffraction effect in free space.

In theory, anti-diffracting light beams are generally obtained by the Fourier transformation of their Fourier spectrums [1-4]. The process of Fourier transformation can be accomplished by an objective lens in Fig. S1 [5, 6]. For example, a quasi-Airy beam can be obtained by focusing a Gaussian beam modulated using a cubic phase plate [2], while a Bessel beam can be created by the Fourier transformation of a Gaussian beam modulated using a high pass filter [1]. Based on Debye vectorial diffractive theory, the electric and magnetic fields of an anti-diffracting light beam near the focus can be expressed as [7, 8]

$$\mathbf{E} \text{ or } \mathbf{H} = -\frac{iA}{\pi} \int_0^{2\pi} \int_0^{\alpha} \sin\theta \cos^{1/2}\theta T_c l_0(\theta) \mathbf{V} \exp(-i k \mathbf{s} \bullet \boldsymbol{\rho}) d\theta d\varphi \tag{3s}$$

where $\theta$ and $\varphi$ are the convergent angle and azimuthal angle, respectively. $A$ is a normalized constant. $\alpha = \arcsin(NA/n)$, $NA$ is the numerical aperture of the objective lens, and $n$ is the refractive index in the focusing space. The wavenumber $k = 2n\pi/\lambda$, where $\lambda$ is the wavelength of the incident beam. $\boldsymbol{\rho} = (r\cos\phi, r\sin\phi, z)$ denotes the position vector of an arbitrary field point in the focal region. In the spherical polar coordinates, the sign of the unit vector $\mathbf{s}$ along a ray in the focusing space is incorrect, as shown in Fig. 1 of Ref. [7], i.e., $\mathbf{s} = (-\sin\theta\cos\varphi, -\sin\theta\sin\varphi, \cos\theta)$ is used instead of $(\sin\theta\cos\varphi, \sin\theta\sin\varphi, \cos\theta)$. $\mathbf{s}$ with the opposite sign will lead to symmetry between calculated and experimental results about the focal plane. $T_c$ is the Fourier spectrum of the desired anti-diffracting light beam. $l_0(\theta)$ is the electric amplitude of the incident Gaussian beam, which can be expressed as

$$l_0(\theta) = \exp[-(\beta_0 \frac{\sin\theta}{\sin\alpha})^2] \tag{4s}$$

where $\beta_0$ is the ratio of the pupil radius to the incident beam waist.

In Eq. (3s), **V** represents the propagating unit vector of the incident beam right after having passed through the lens. Here, we take a linearly polarized beam as an example to generate an anti-diffracting light beam. Thus, the electric vector $\mathbf{V}_e$ and magnetic vector $\mathbf{V}_m$ can be written as [7]

$$\mathbf{V}_e = \begin{bmatrix} \cos\theta + (1-\cos\theta)\sin^2\varphi \\ -(1-\cos\theta)\sin\varphi\cos\varphi \\ \sin\theta\cos\varphi \end{bmatrix}; \quad \mathbf{V}_m = \begin{bmatrix} -(1-\cos\theta)\sin\varphi\cos\varphi \\ 1-(1-\cos\theta)\sin^2\varphi \\ \sin\theta\sin\varphi \end{bmatrix} \quad (5s)$$

Eventually, the light intensity distribution of an anti-diffracting light beam can be obtained using $I = |\mathbf{E}|^2$. Moreover, the time-averaged Poynting vector, namely, the energy flux, can be obtained using [9]

$$\langle \mathbf{S} \rangle = \frac{c}{4\pi} \text{Re}(\mathbf{E} \times \mathbf{H}^*) \quad (6s)$$

where $c$ is the velocity of light in a vacuum, and the asterisk denotes the operation of complex conjugation. According to Eq. (6s), one can flexibly explore the energy process of anti-diffracting light beam during propagation in free space.

Supposed $P_1(x,y,z)$ and $P_2(x,y,-z)$ are two symmetrical points on the anti-diffracting light beam, the electric and magnetic fields of which can be expressed as [7]

$$\begin{array}{ll} E_x(x,y,-z) = -E_x^*(x,y,z) & H_x(x,y,-z) = -H_x^*(x,y,z) \\ E_y(x,y,-z) = -E_y^*(x,y,z); & H_y(x,y,-z) = -H_y^*(x,y,z) \\ E_z(x,y,-z) = E_z^*(x,y,z) & H_z(x,y,-z) = H_z^*(x,y,z) \end{array} \quad (7s)$$

By substituting Eq. (7s) into Eq. (6s), the relationship of transverse energy fluxes between $P_1(x,y,z)$ and $P_2(x,y,-z)$ can be expressed as

$$\langle \mathbf{S} \rangle_{tP_1} = -\langle \mathbf{S} \rangle_{tP_2} \quad (8s)$$

which implies that the light beam experiences two inverse energy processes that transfer the energy from $\langle \mathbf{S} \rangle_{tP_2}$ to $\langle \mathbf{S} \rangle_{tP_1}$ during propagating in free space. Without loss of generality, $\langle \mathbf{S} \rangle_{tP_2}$ and $\langle \mathbf{S} \rangle_{tP_1}$ are called as energy charge and energy discharge, respectively.

Energy charge and discharge compose an entire energy oscillation, which is a directional energy flux that confines the energy of anti-diffracting light beam into an interaction between mainlobe and sidelobes when propagating in free space. Thus, the light beam would not diverge freely like that of Gaussian beam in free space. If only energy oscillation occurs, anti-diffracting light beam can preserve its shape without divergence in free space. Even when encountering an obstacle, the mainlobe can carry out self-healing with the power from the sidelobes [10]. This is the reason why anti-diffracting light beam is naturally composed with mainlobe and sidelobes. Moreover, since all anti-diffracting light beams can be created by focusing their corresponding Fourier spectrums in Fig. S1, energy oscillation is therefore a general property shared by all anti-diffracting light beams.

## A. Energy oscillation mechanism of an Airy beam

Although energy oscillation is a general property shared by all anti-diffracting light beams, different light beam exhibits different form of energy oscillation, which is mainly determined by the Fourier spectrum $T_c$. For example, cubic phase is the Fourier spectrums of Airy beam, the transmittance of which can be written as [2, 3]

$$T_c = \exp\left\{i\eta \frac{k\sin^\sigma \theta}{\sin^\sigma \alpha}\left[\sin^3(\varphi+\phi_0)+\cos^3(\varphi+\phi_0)\right]\right\} \tag{9s}$$

where $\eta$, $\sigma$ are the parameters that control the period and phase distribution of the cubic phase plate, respectively. Typically, if $\sigma = 3$, $T_c$ denotes a standard cubic phase, and the whole phase can be rotated by an angle $\phi_0$. Accordingly, a quasi-Airy beam generated using the cubic phase of $-\eta$ can simply be obtained by rotating the cubic phase of $\eta$ with $\phi_0 = \pi$. That is, light beams with $\pm\eta$ are symmetrical about the optical axis, thereby leading to

$$\langle \mathbf{S}_\eta \rangle_{tP_1} = -\langle \mathbf{S}_{-\eta} \rangle_{tP_1} \tag{10s}$$

where $\langle \mathbf{S}_\eta \rangle_{tP_1}$ and $\langle \mathbf{S}_{-\eta} \rangle_{tP_1}$ are the transverse energy fluxes of point $P_1(x,y,z)$ for $\eta$ and $-\eta$, respectively.

From Eq. (8s, 10s), the transverse energy flux relationship between quasi-Airy beams with $\pm\eta$ can be simplified to

$$\langle \mathbf{S}_\eta \rangle_{tP_1} = \langle \mathbf{S}_{-\eta} \rangle_{tP_2} = -\langle \mathbf{S}_\eta \rangle_{tP_2} = -\langle \mathbf{S}_{-\eta} \rangle_{tP_1} \tag{11s}$$

which denotes that the energy charge $\langle \mathbf{S}_\eta \rangle_{tP_2}$ is equal with the energy discharge $\langle \mathbf{S}_{-\eta} \rangle_{tP_1}$, while the energy discharge $\langle \mathbf{S}_\eta \rangle_{tP_1}$ is equal with the energy charge $\langle \mathbf{S}_{-\eta} \rangle_{tP_2}$. Thus, quasi-Airy beams with $\pm\eta$ can be considered as a pair of mutually complementary modes in free space.

**Energy interaction between mainlobe and sidelobes**

In the following simulations and experiments, $NA=0.095$, $n=1$, and $\beta_0 = 1$. The unit of length in all figures is the wavelength $\lambda$, and the light intensity is normalized to the unit value.

The quasi-Airy beam in Fig. S3(a), generated by the cubic phase with the parameters $\eta = 5$, $\sigma = 3$ and $\phi_0 = 0.75\pi$ in Fig. S3(b), represents one entire energy oscillation in free space, which is composed of energy charge when $z<0$ and discharge when $z>0$. Energy charge and discharge imply two different energy processes. In the process of energy charge when $z<0$, the energy is transported from the mainlobe to the sidelobes, as shown in Fig. S4(a, b). Thus, the mainlobe is the energy source of sidelobes, and the energy of mainlobe tends to be stored in the sidelobes instead of diverging like that of a Gaussian beam. In contrast to energy charge, the light beam when $z>0$ experiences a totally different energy process, namely, energy discharge. Due to the inverse energy flux in Fig. S4(c, d), the energy is transported from the sidelobes to the mainlobe. In this case, the sidelobes are the energy sources of mainlobe, while the mainlobe becomes an energy consumer. During propagation when $z>0$, the mainlobe endures an energy loss caused by the diffraction effect. However, this can be replenished by the power of sidelobes, thus the light beam can remain anti-diffracting. Even when encountering an obstacle, the light beam can carry out self-healing through the power of sidelobes [10].

Due to the energy oscillation mechanism, the power of quasi-Airy beam is confined to the interplay between the mainlobe and sidelobes. Therefore, the light beam can propagate without significant divergence in free space. However, this confinement cannot be limitless. Finite power in free space can support only a finite energy charge when $z<0$, thus leading to only a finite energy discharge when $z>0$. When the power stored in the sidelobes is exhausted, the light beam when $z>0$ can no longer maintain its shape, and the diffraction effect eventually dominates. Consequently, the light beam can propagate only a finite anti-diffracting distance in free space. In addition, the power of energy charge is equal to that of energy discharge because of energy conversation, thereby giving rise to an equivalent non-diffractive distance when $z<0$ and $z>0$. For this reason, a quasi-Airy beam possesses a symmetrical trajectory, as shown in Fig. S3(a).

In the whole energy oscillation process, the $z=0$ plane is the inflection plane, in which the role of the mainlobe changes from energy source to energy consumer. Specifically, when $z<0$, the sidelobes can receive power from the mainlobe continually during the energy charge process. However, once the sidelobes attain maximum energy capacity, energy charge can no longer be conducted, and the sidelobes begin providing power to the mainlobe when $z>0$. Thus, the sidelobes have maximum light intensity in the $z=0$ plane. Suppose that $I_s$ is the light intensity of sidelobe A in the $z=0$ plane, as shown in Fig. S4(a).

In principle, larger $I_s$ indicates stronger energy oscillation, thereby leading to a longer anti-diffracting distance of light beam in free space. That is, if $I_s$ can be adjusted, the strength of the energy oscillation, along with the non-diffractive distance, can be controlled accordingly.

**Adjusting the strength of Energy oscillation**

Two methods are proposed to control the strength of energy oscillation via $I_s$. One is to adjust the period of cubic phase $\eta$ while $\sigma = 3$; the other is to manipulate the cubic phase distribution via $\sigma$ while keeping the period $\eta$ unchanged. For the first one, quasi-Airy beams with different values of $\eta$ are generated in Fig. S5, where (a, f) $\eta = 3$ and (b, i) $\eta = 5$. Figures S5(e, h) depict their corresponding cubic phase plates. The theoretical results show that $I_s$ increases with the period of the cubic phase $\eta$, where $I_s$=0.213 for $\eta = 3$ [Fig. S5(f)] and $I_s$=0.239 for $\eta = 5$ [Fig. S5(i)]. Therefore, a longer non-diffractive distance can be achieved when $\eta$ is larger; the FWHMs of Fig. S5(a, b) are 1515.6λ and 2560λ, respectively. By adopting the same experimental setup as in Fig. 1 [see Main text], quasi-Airy beams with $\eta = 3$ [Fig. S5(c)] and $\eta = 5$ [Fig. S5(d)] are created in the condition of *NA*=0.095. The FWHMs are 1522.1λ and 2544.9λ, respectively, which are coincident with those of Fig. S5(a, b) in theory.

As for the second one, Figure S6 shows the light intensity of quasi-Airy beams with (a, f) $\sigma = 2.6$ and (b, i) $\sigma = 3.6$ when $\eta = 5$. As shown in Fig. S6(f, i), $I_s$ increases as $\sigma$ decreases, with $I_s$=0.355 [Fig. S6(f)] and $I_s$=0.239 [Fig. S6(i)]. Likewise, the non-diffractive distance of $\sigma = 2.6$ is much longer than that of $\sigma = 3.6$. In the experiment, the quasi-Airy beams shown in Fig. S6(a, f) and (b, i) are created by coding the phase of Fig. S6(e, h) in the SLM shown in Fig. 1; see Main text. The FWHM values for both beams are 3077.6λ and 1419.3λ, respectively, which fit well with those of $\sigma = 2.6$ (3084λ) and 3.6 (1411λ) in theory. Although both methods mentioned above are capable of adjusting the energy strength via the light intensity of sidelobe $I_s$, none can strengthen the energy oscillation without limitation, which makes the ideal Airy beam impossible in practice. However, this new mechanism still offers a new possibility for light beams to break through the limitation of finite non-diffractive distance in free space.

## B. Energy oscillation mechanism of a Bessel beam

Figure S7 presents the light intensity and energy flux of a Bessel beam, which is generated by focusing a linearly polarized Gaussian beam with the modulation of high-pass pupil filter shown in Fig. S7(c) [1]. Here, the transmission of this pupil filter can be expressed as

$$T_h = \begin{cases} 1 & r/R \geq 0.8 \\ 0 & 0 \leq r/R < 0.8 \end{cases} \qquad (12s)$$

where $r$ and $R$ are the radii of the inner and outer ring of the high-pass pupil filter, respectively. By substituting $T_c$ for $T_h$ in Eq. (3s), the light intensity and energy flux of the Bessel beam can be calculated using Debye vectorial diffractive theory. In the following simulations, $NA$=0.095, $n$=1, and $\beta_0 = 1$. The unit of length in all figures is the wavelength λ, and the light intensity is normalized to the unit value.

As shown in Fig. S7(c, d), a Bessel beam can be divided into two parts according to the energy flux: energy charge when z<0 and energy discharge when z>0. For the light beam when z<0, the energy is transported from the sidelobes to the central mainlobe radially. Thus, the sidelobes are the energy source of the mainlobe. Even when encountering an obstacle, the light beam can carry out self-healing [11]. In contrast to the light beam when z<0, the light beam when z>0 experiences an inverse energy process. As shown in Fig. S7(d), the energy is transported from the mainlobe to the sidelobes radially. In this case, the mainlobe turns into an energy consumer, while the sidelobes serve as the energy source of the mainlobe. Via this energy interaction between the mainlobe and sidelobes, the Bessel beam can preserve its shape without significance divergence in free space. Similar to the quasi-Airy beam in Fig. S4, finite power in free space can support only a finite energy oscillation, thus leading to a finite anti-diffracting distance of Bessel beam.

# 3. Derivation of 'optical pen'

Figure S8 presents a schematic of the focusing system. A collimate incident vector beam propagating along the +z-axis goes through the pupil filter P before being focused by a lens obeying the sine condition. $\Omega$ is the focal sphere, with its center at $O$ and a radius $f$, namely, the focal length of the lens. $O_1$ is an arbitrary point in the focal region of the lens. In principle, only one focus is located at the focal point $O$, where constructive interference can occur only because of the equivalent optical paths of the light beams between the points in $\Omega$ and point $O$. Likewise, if constructive interference does not occur at point $O$ but at arbitrary point $O_1$, the focus is generated at point $O_1$ instead of at $O$. To this end, the optical path difference (OPD) for the light beams between the points in $\Omega$ and $O_1$ must be compensated by the phase of pupil filter P.

As shown in Fig. S8, the OPD for the light beams between the points in $\Omega$ and $O_1$ can be simplified to $L_2 - L_1$, where $L_1$, $L_2$ are the light paths $AO_1$ and $BO_1$, respectively. Here, in the cylindrical coordinate system, the points A, B, O and $O_1$ can be expressed as $(f\sin\theta\cos\varphi, f\sin\theta\sin\varphi, -f\cos\theta)$, $(0,0,-f)$, $(0,0,0)$ and $(\rho\cos\varphi_s, \rho\sin\varphi_s, z)$, respectively. The light paths $L_1$ and $L_2$ can be described as

$$L_1 = \sqrt{(f\sin\theta\cos\varphi - \rho\cos\varphi_s)^2 + (f\sin\theta\sin\varphi - \rho\sin\varphi_s)^2 + (f\cos\theta + z)^2}$$
$$= \sqrt{f^2 + \rho^2 + z^2 - 2f\rho\sin\theta\cos(\varphi - \varphi_s) + 2fz\cos\theta} \quad (13s)$$

$$L_2 = \sqrt{(\rho\cos\varphi_s)^2 + (\rho\sin\varphi_s)^2 + (-f-z)^2}$$
$$= \sqrt{\rho^2 + f^2 + z^2 + 2fz} \quad (14s)$$

The OPD for the light beams between the points in $\Omega$ and $O_1$ can be calculated as

$$\Delta s = L_2 - L_1 = \frac{2\rho\sin\theta\cos(\varphi - \varphi_s) + 2z(1 - \cos\theta)}{\sqrt{\eta_\rho^2 + 1 + \eta_z^2 + 2\eta_z} + \sqrt{1 + \eta_\rho^2 + \eta_z^2 - 2\eta_\rho\sin\theta\cos(\varphi - \varphi_s) + 2\eta_z\cos\theta}} \quad (15s)$$

where $\theta$ and $\varphi$ are the convergent angle and azimuthal angle, respectively. $\eta_z = z/f$, and $\eta_\rho = \rho/f$. Since $O_1$ is an arbitrary point in the vicinity of the focus, $\rho, z \ll f$. That is, $\eta_\rho, \eta_z \approx 0$. Finally, Eq. (15s) can be simplified to

$$\Delta s = \rho\sin\theta\cos(\varphi - \varphi_s) + z(1 - \cos\theta) \quad (16s)$$

To generate a focus at $O_1$, $\Delta s$ must be compensated by the pupil filter P. Thus, $\Delta s_p = -\Delta s$, where $\Delta s_p$ is the OPD induced by the pupil filter P. According to the relationship between the phase and OPD, the phase of pupil filter P is $\psi = k\Delta s_p$, where $k = 2\pi n / \lambda$ is the wavenumber and $n$ is the refractive index in the focusing space. Consequently, the transmission of pupil filter P for one single focus can be expressed as $T = \exp(i\psi)$. If there are multiple focuses in the focal region, the transmittance of pupil filter P can be further transformed into

$$T = \sum_{i=1}^{N} s_i \exp(i\psi_i) \tag{17s}$$

where $N$ is the number of foci and $i$ denotes the $i$-th focus. As shown in Eq. (17s), the pupil filter P requires both amplitude and phase modulation for the incident light beam. In practice, amplitude modulation always leads to low light transformation efficiency and is difficult to implement. However, this problem can easily be solved by extracting only the phase of pupil filter P in Eq. (17s). Finally, the phase-only pupil filter P can be obtained using

$$T = \exp\left\{iPhase\left\{\sum_{i=1}^{N} s_i \exp[i(\psi_i + \delta_i)]\right\}\right\} \tag{18s}$$

Here, we refer to this pupil filter as the 'optical pen'. For the sake of simplicity, this 'optical pen' can also be expressed as

$$T = \exp\left\{iPhase[\sum_{i=1}^{N} PF(s_i, x_i, y_i, z_i, \delta_i)]\right\} \tag{19s}$$

where $x_i, y_i, z_i$ are the positions of the $i$-th focus in the focal region corresponding to $\rho, \varphi_s, z$ in Eq. (16s). $s_i$ and $\delta_i$ are two weight factors that are responsible for adjusting the amplitude and phase of the $i$-th focus.

## Manipulation of light field via "optical pen"

Note that the pupil filter P developed by the OPD compensation is valid for an arbitrary incident vector beam. Here, we take only a linearly polarized beam as example to verify the function of the 'optical pen'. Based on Debye vectorial diffractive theory, the light intensity in the vicinity of the focus can be obtained using Eq. (3s). In the following simulations and experiments, $NA=0.8$, $n=1$, and $\beta_0 = 1$. The unit of length in all figures is the wavelength λ, and the light intensity is normalized to the unit value.

In the case of $N = 1$, only one focus is obtained in the focal region, the position of which can be adjusted in three dimensions using the 'optical pen'. As shown in Fig. S9, foci with different positions in the y-z plane are realized with the different phases of pupil filter P [namely, Fig. S9(b, d, f)], the parameters of which are $N = 1$, $s_1 = 1$, $x_1 = 1$, $\delta = 0$; (a, b) $y_1 = -5, z_1 = 10$, (c, d) $y_1 = 0, z_1 = 0$, and (e, f)

$y_1 = 5, z_1 = -10$, respectively. By comparing with the original focus in Fig. S9(c), the shape of the focus remains invariant, while the position can be adjusted freely in the focal region.

For an 'optical pen', the focal region can be considered as a drawing board, on which an arbitrary pattern can be realized by precisely controlling the number and position of foci and their corresponding weight factors in Eq. (19s). The size of the 'optical pen' is determined by the focus of the incident linearly polarized beam, which is relevant to the *NA*. For one particular light pattern, the number and position of foci determine the shape of light pattern, which can be manipulated using the parameters $N, x_i, y_i, z_i$, respectively. Once the shape is confirmed, the amplitude and phase of each focus must be adjusted by $s_i$ and $\delta_i$ so that the desired light pattern can be realized in the focal region.

Regarding the same light pattern, the number and position of foci are identical, which is determined only by its shape. However, the weight factors $s_i$ and $\delta_i$ have countless possibilities. As shown in Fig. S10, a 4×4 focal array in the x-y plane can be realized with the different phases shown in Fig. S10(d-f), respectively. Figures S10(d-f) can be obtained by multiplying two phases, yielding a 1×4 focal array along the x- and y-axis, respectively. Thus, $N = 4$, $z_i = 0$, and $\delta_i = 0$. For the 1×4 foci along the x-axis, $y_i = 0$, with the foci located at $x_1 = -3$, $x_2 = 3$, $x_3 = -9$ and $x_4 = 9$. For the 1×4 foci along the y-axis, $x_i = 0$, with the foci located at $y_1 = -3$, $y_2 = 3$, $y_3 = -9$ and $y_4 = 9$. The distance between each focus along the x- and y-axis is $6\lambda$. The only difference between the phases for the same 1×4 foci is the weight factor $s_i$. For example, three pupil filters for the 1×4 focal array along the x- and y-axis are obtained, with (i) $s_1 = 1.05, s_2 = 0.7, s_3 = 0.92, s_4 = -0.28$; (ii) $s_1 = s_2 = -0.885, s_3 = s_4 = 1$; and (iii) $s_1 = -1.05, s_2 = 0.7$, $s_3 = 0.95, s_4 = 0.33$, which are denoted as $l_{x1} \sim l_{x3}$ for the x-axis and $l_{y1} \sim l_{y3}$ for the y-axis. Finally, the phases can be obtained as follows: (d) $l_{x1} \times l_{y1}$; (e) $l_{x2} \times l_{y2}$; (f) $l_{x3} \times l_{y2}$, respectively. Clearly, all phases in Fig. S10(d-f) are different from each other, but all generate an identical 4×4 focal array in the focal plane. Among these, the most typical one is the phase in Fig. S10(e), which is a common Dammann grating and was also obtained in Ref. [12]. In other words, the 'optical pen' in Eq. (19s) represents all possible phases, with only the weight factors being different; the Dammann grating is only one of the special solutions. Thus, more phases can be obtained by adjusting the weight factor $s_i$.

When generating a light pattern in the focal region, the shape is determined by the number and position of foci, while the amplitude and phase of each focus ensure the quality of the whole pattern. By controlling the number and position along with the weight factors $s_i$ and $\delta_i$, more complex focal patterns can be realized. As shown in Fig. S11(a, b), "NANO" and "OPT" are created in the focal plane with the phases in Fig. S11(c, d). In addition, a three-dimensional focal pattern can also be realized using the 'optical pen'. As shown in Fig. S12(a), "OPT" is obtained in the $z = -20, 0, 20$ planes simultaneously with the phase shown in Fig. S12(b). All parameters for the above phases can be found in the Appendix. Once again, the phases for the generation of focal patterns in Fig. S11,12 are not unique, but all can be obtained using $s_i$ and $\delta_i$ in Eq. (19s).

## The significance of 'optical pen'

The generation of an ultralong anti-diffracting light beam in free space always leads to precise manipulation of the number, position, amplitude and phase of foci in the focal region. However, this cannot be achieved using past techniques [12-17]. The 'optical pen' is therefore developed to solve this problem. As a versatile optical tool, the 'optical pen' possesses an explicit form (Eq. (19s)), which can be used to unify the relationship between the focal pattern and the phase in the entrance plane. By adjusting the parameters of the 'optical pen', one can adjust the number, position, amplitude and phase of foci at will in the focal region so that an arbitrary focal pattern can be realized in free space. This advantage makes the 'optical pen' a perfect optical tool for the creation of ultralong anti-diffracting light beams in free space.

## 4. Creating UAD light beam in the condition of *NA*=0.8

Figure S13 shows UAD light beams with 0, 1, 2, and 3 energy oscillations in the y-z plane, generated by focusing a linearly polarized Gaussian beam with *NA*=0.8. The corresponding phases are shown in Fig. S13(e-h), respectively, and the parameters can be found in Table S1. Notably, the anti-diffracting distances of the UAD light beams for zero and one energy oscillation are only $2.34\lambda$ and $31.96\lambda$, respectively. However, using the multiple energy oscillation mechanism, UAD light beams can easily be achieved by simply increasing the number of energy oscillations via the modulation of the 'optical pen'. As shown in Fig. S13(c, d), the anti-diffracting distances of two and three energy oscillations are $70.56\lambda$ and $99.50\lambda$, respectively.

The multiple energy oscillation mechanism can lead to a peculiar energy flux at the switch point between adjacent energy oscillations, as indicated by the cross-shaped sidelobes in Fig. S13(B). The bottom sidelobe corresponds to an energy discharge process similar to that at point A in the initial energy oscillation, as shown in Fig. S13(A). Although the energy discharge at Point B will soon be over, the upper sidelobes can provide an additional energy recharge so that the light beam can propagate further. That is, at Point B, the light beam experiences not only an energy discharge in the initial energy oscillation but also an energy charge similar to that at point C in Fig. S13(C) in the second energy oscillation. Thus, the UAD light beam can maintain its shape without significant divergence over a super-long range.

# Appendix

**Phase of the pupil filter for "NANO":**

1. Parameters for 'O'

    $d = -9$ ; $N = 16$;
    $\varphi_i = 2\pi(i-1)/N, i = 1,2,3...N$
    $\begin{cases} x_i = d\cos\phi_i \\ y_i = d\sin\phi_i \end{cases}$
    $l_O = \sum_{i=1}^{N=16} PF(s_i, x_i, y_i, 0).$
    $s_{1-4} = s_{6,7} = s_{9-12} = s_{16} = 1$; $s_5 = 0.97$; $s_8 = 1.05$; $s_{13} = 1.05$; $s_{14} = 1.15$; $s_{15} = 0.98$

2. Parameters for 'N'

    $d_2 = -7.5; d_3 = -10;$
    $l_{N1} = PF(1, d_2, -d_3, 0) + PF(0.97, -d_2, -d_3, 0) + PF(-1, d_2, -d_3/2, 0) + PF(0.95, d_2/2, -d_3/2, 0) +$
    $\quad PF(-0.97, -d_2, -d_3/2, 0) + PF(-0.8, d_2, 0, 0) + PF(1.05, 0, 0, 0) + PF(-0.8, -d_2, 0, 0) +$
    $\quad PF(-0.85, d_2, d_3/2, 0) + PF(0.95, -d_2/2, d_3/2, 0) + PF(-0.88, -d_2, d_3/2, 0) + PF(0.88, d_2, d_3, 0) +$
    $\quad PF(1, -d_2, d_3, 0) + PF(-0.1, 0, 15, 0) + PF(-0.1, -d_2/2, -d_3, 0).$
    $l_{N2} = PF(0.98, d_2, -d_3, 0) + PF(0.93, -d_2, -d_3, 0) + PF(-0.88, d_2, -d_3/2, 0) + PF(0.95, d_2/2, -d_3/2, 0) +$
    $\quad PF(-0.85, -d_2, -d_3/2, 0) + PF(-0.8, d_2, 0, 0) + PF(1.05, 0, 0, 0) + PF(-0.8, -d_2, 0, 0) +$
    $\quad PF(-0.91, d_2, d_3/2, 0) + PF(0.943, -d_2/2, d_3/2, 0) + PF(-0.93, -d_2, d_3/2, 0) + PF(0.9, d_2, d_3, 0) +$
    $\quad PF(1, -d_2, d_3, 0) + PF(-0.2, -d_2/2, d_3, 0).$

3. Parameters for 'A'

    $d_4 = -10; d_5 = -7.5;$
    $l_A = PF(0.83, 0, -d_4, 0) + PF(-0.9, d_4/4, -d_4/2, 0) + PF(-0.9, -d_4/4, -d_4/2, 0) +$
    $\quad PF(0.9, d_4/2, 0, 0) + PF(0.9, -d_4/2, 0, 0) + PF(0.8, d_5, d_4/2, 0) + PF(0.89, -d_5, d_4/2, 0) +$
    $\quad PF(-0.82, d_4/4, d_4/2, 0) + PF(-0.83, -d_4/4, d_4/2, 0) + PF(0.8, d_4, d_4, 0) + PF(0.83, -d_4, d_4, 0).$

4. Final phase

    $d_6 = 15$
    $\psi_{pf} = Phase[PF(1.3, -d_6, -d_6, 0)l_{N1} + PF(1.3, -d_6, d_6, 0)l_{N2} + PF(1, d_6, d_6, 0)l_A + PF(0.9, d_6, -d_6, 0)l_O]$
    $l_{pf} = \exp(i\psi_{pf})$

**Phase of the pupil filter for "OPT":**

1. Parameters for 'O'

    $d = -9$ ; $N = 16$;
    $\varphi_i = 2\pi(i-1)/N, i = 1,2,3...N$
    $\begin{cases} x_i = d\cos\phi_i \\ y_i = d\sin\phi_i \end{cases}$
    $l_O = \sum_{i=1}^{N=16} PF(s_i, x_i, y_i, 0).$
    $s_{1,3} = s_{6,7} = 1$; $s_{2,11} = 1.03$; $s_4 = 1.01$; $s_5 = 1.15$; $s_8 = 1.1$;
    $s_9 = 1.08$; $s_{10} = 1.05$; $s_{13} = 1.24$; $s_{12,14-16} = 1.02$

2. Parameters for 'P'

$$d_i = (i-4) \times 3$$
$$l_P = \sum_{i=1}^{7} PF(s_i,0,d_i,0) + PF(s_8,3,0,0) + PF(s_9,6,0,0) + PF(s_{10},3,9,0) + PF(s_{11},6,9,0) +$$
$$PF(s_{12},8,6.75,0) + PF(s_{13},8,2.25,0) + PF(s_{14},3,-9,0)$$
$$s_1 = 1.125; s_2 = 1.12; s_{3,5} = -1.205; s_4 = 1.505; s_6 = 0.935; s_{7,13} = 1.1;$$
$$s_8 = 1.6; s_9 = 1.4; s_{10} = 1.45; s_{11,12} = 1.2; s_{14} = 0.3;$$

3. Parameters for 'T'
$$d_j = (j-4) \times 3; D_j = (j-10) \times 3$$
$$l_T = \sum_{i=1}^{7} PF(s_i,0,d_i,0) + \sum_{i=8}^{9} PF(s_i,D_i,9,0) + \sum_{i=10}^{11} PF(s_i,D_{i+1},9,0)$$
$$s_1 = 1.4; s_2 = -1.26; s_3 = 1.3; s_4 = 1.1; s_5 = 1.25; s_6 = -1.35;$$
$$s_7 = 1.55; s_8 = -0.95; s_{9,10} = 1.46; s_{11} = -1.15;$$

4. Final phase
$$\psi_{pf} = Phase[PF(1.45,-20,0,0)l_O + PF(1.1,-3,0,0)l_P + PF(1.1,20,0,0)l_T]$$
$$l_{pf} = \exp(i\psi_{pf})$$

**Phase of the pupil filter for "OPT" in different z planes:**

1. Parameters for 'O'
$$d = -9; N = 16;$$
$$\varphi_i = 2\pi(i-1)/N, i = 1,2,3...N$$
$$\begin{cases} x_i = d\cos\phi_i \\ y_i = d\sin\phi_i \end{cases}$$
$$l_O = \sum_{i=1}^{N=16} PF(s_i,x_i,y_i,0).$$
$$s_1 = 0.95; s_{13} = 1.2; s_{2,4,6-8,14-16} = 1; s_{3,5,9-11} = 1.1$$

2. Parameters for 'P'
$$d_i = (i-4) \times 3$$
$$l_P = \sum_{i=1}^{7} PF(s_i,0,d_i,0) + PF(s_8,3,0,0) + PF(s_9,6,0,0) + PF(s_{10},3,9,0) + PF(s_{11},6,9,0) +$$
$$PF(s_{12},8,6.75,0) + PF(s_{13},8,2.25,0)$$
$$s_{1,7} = 0.8; s_{2,10,11} = 1.3; s_3 = -1.3; s_4 = 1.6; s_5 = -1.4; s_{6,9} = 1.2; s_8 = 1.4; s_{12} = 1.15; s_{13} = 1;$$

3. Parameters for 'T'
$$d_i = (i-4) \times 3, D_i = (i-10) \times 3$$
$$l_T = \sum_{i=1}^{7} PF(s_i,0,d_i,0) + \sum_{i=8}^{9} PF(s_i,D_i,9,0) + \sum_{i=10}^{11} PF(s_i,D_{i+1},9,0)$$
$$s_1 = 1.45; s_2 = -1.35; s_3 = 1.35; s_4 = 0.8; s_5 = 1.23; s_6 = -1.25; s_7 = 1.4; s_8 = -1; s_{9,10} = 1.3; s_{11} = -0.95.$$

4. Final phase
$$\psi_{pf} = Phase[PF(1.25,-12,-12,-20)l_O + PF(0.98,0,0,0)l_P + PF(1,12,12,20)l_T].$$
$$l_{pf} = \exp(i\psi_{pf})$$

# Figure

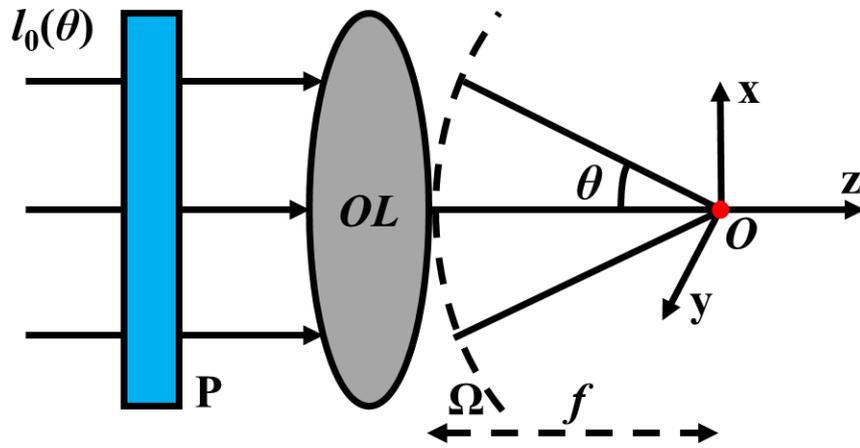

**Figure S1** Schematic of the focusing system. $\Omega$ is the focal sphere, with its center at $O$ and a radius $f$, namely, the focal length of the lens. P is the pupil filter in the wavefront of the lens (OL). $l_0(\theta)$ denotes the electric amplitude of incident Gaussian beams.

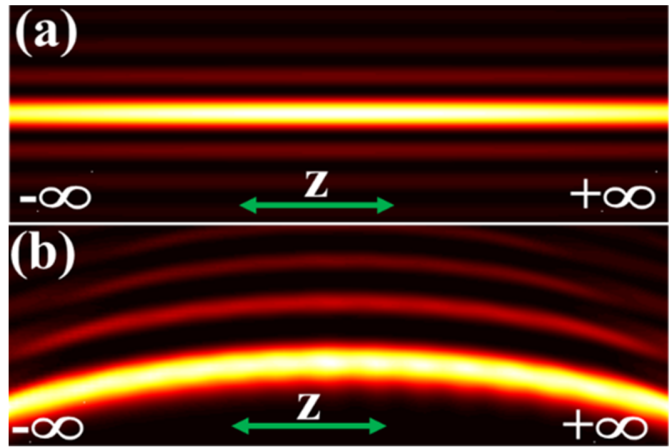

**Figure S2** The light intensity of ideal Bessel beam (a) and Airy beam (b).

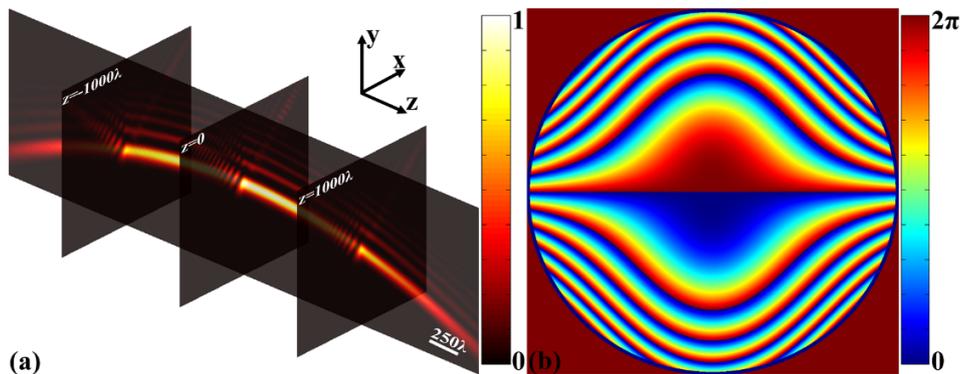

**Figure S3** Quasi-Airy beam (a) generated by the cubic phase plate (b), the parameters of which are $\eta = 5$, $\sigma = 3$ and $\phi_0 = 0.75\pi$.

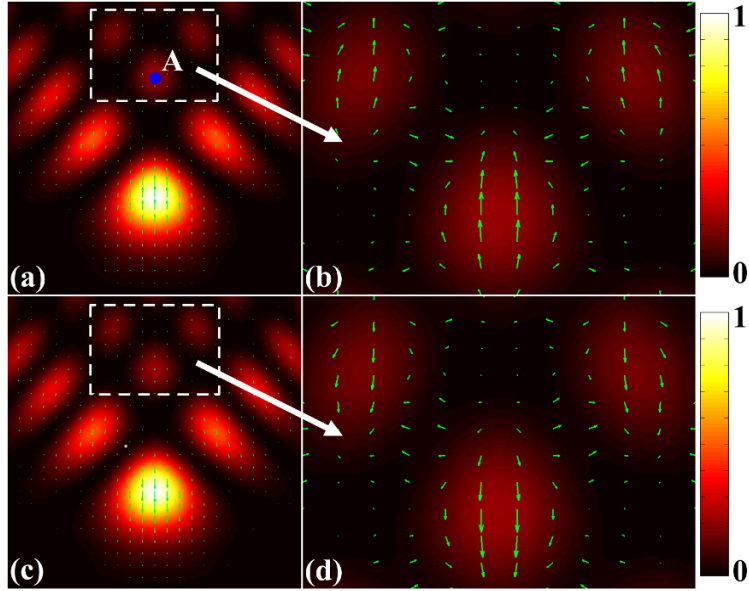

**Figure S4** Light intensity of a quasi-Airy beam in the z=−1000λ (a, b) and z=1000λ (c, d) planes. Their corresponding transverse energy fluxes are indicated by the green arrows. Point A denotes the first sidelobe of the light beam.

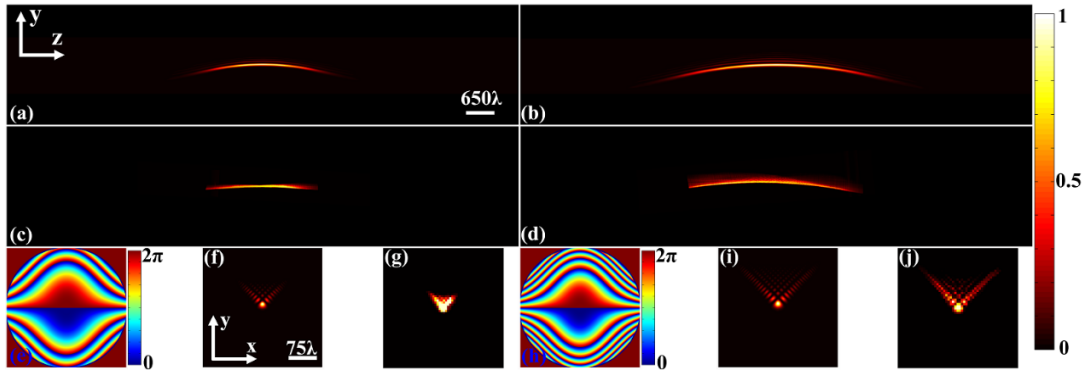

**Figure S5** Quasi-Airy beams generated by the cubic phases with $\eta = 3$ (e) and $\eta = 5$ (h), with $NA$=0.095 and $\sigma = 3$. The light intensities in the y-z and z=0 planes are shown in (a, b) and (f, i), respectively. (c, d) and (g, j) present their corresponding experimental results. The FWHMs of (a, b) are theoretically approximately 1515.6λ and 2560λ, while their experimental results are (c) 1522.1λ and (d) 2544.9λ, respectively.

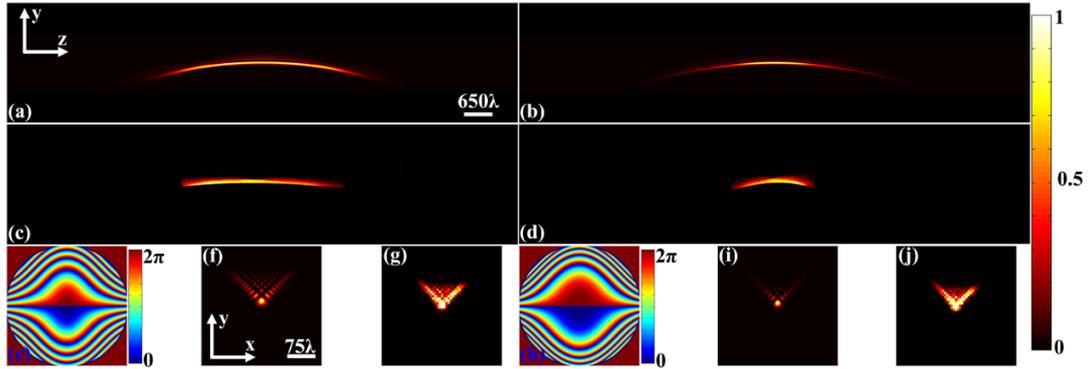

**Figure S6** Quasi-Airy beams generated by the cubic phases with $\sigma = 2.6$ (e) and $\sigma = 3.6$ (h), with $NA=0.095$ and $\eta = 5$. The light intensities in the y-z and z=0 planes are shown in (a, b) and (f, i), respectively. (c, d) and (g, j) present their corresponding experimental results. The FWHMs of (a, b) are theoretically approximately $3084\lambda$ and $1411\lambda$, while their experimental results are (c) $3077.6\lambda$ and (d) $1419.3\lambda$, respectively.

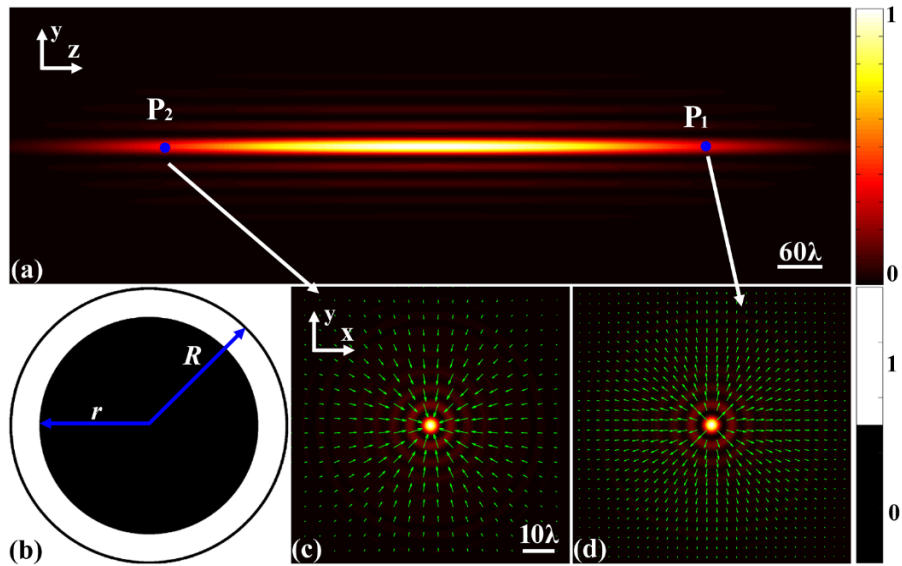

**Figure S7** Bessel beam (a) generated using the high-pass pupil filter with a ratio of the inner to outer ring radius $r/R=0.8$ (b); (c, d) present the light intensities and energy fluxes (green arrows) at points $P_2$, $P_1$ in the $z=-300\lambda$, $300\lambda$ planes, respectively.

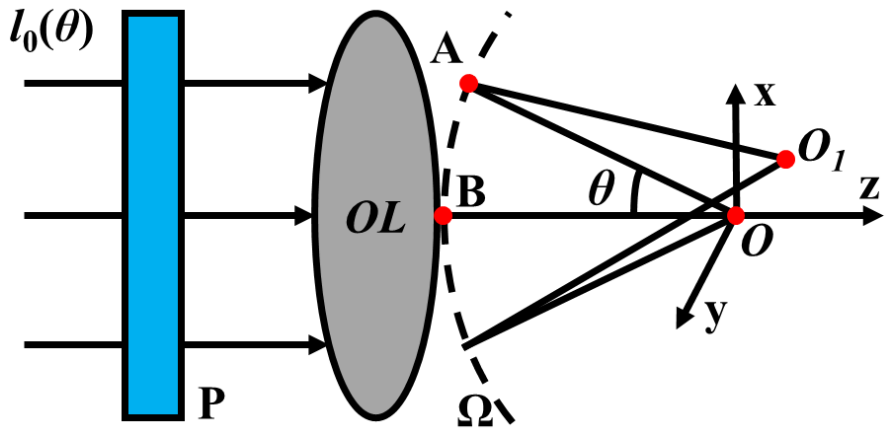

**Figure S8** Schematic of the focusing system. $\Omega$ is the focal sphere, with its center at $O$ and a radius $f$. A, B are the off- and on-axis points in $\Omega$. $O_1$ is an arbitrary point in the focal region. P is the pupil filter in the wavefront of the lens (OL). $l_0(\theta)$ denotes the electric amplitude of incident Gaussian beams.

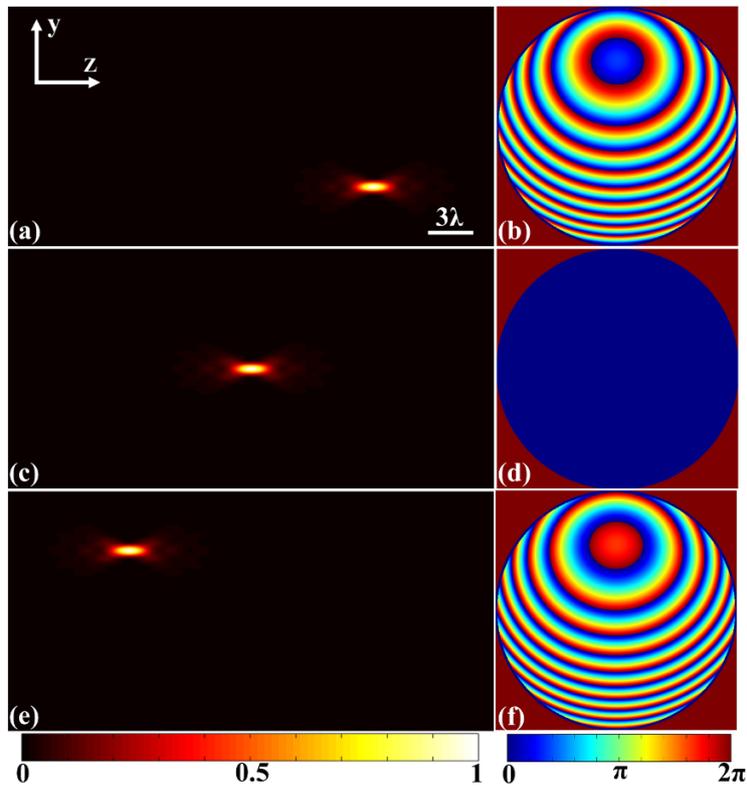

**Figure S9** Shifting focuses in the y-z plane with the phases in (b, d, f), respectively, where $N=1$, $s_1=1$, $\delta=0$, and x=0. (a) y=−5, z=10; (c) y=0, z=0; (e) y=5, z=−10.

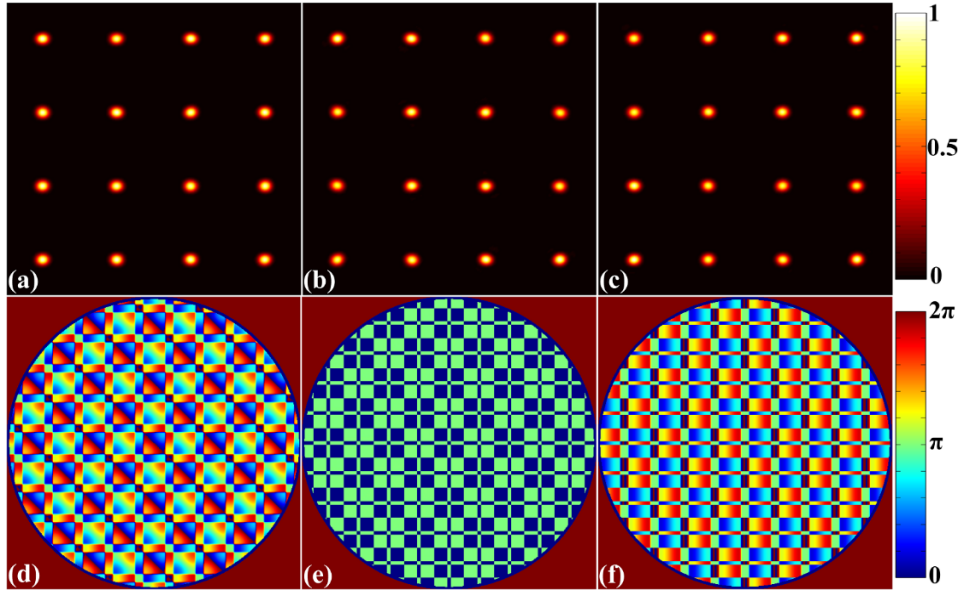

**Figure S10** 4×4 focal arrays generated in the focal plane (a-c); their corresponding phases are shown in (d-e), respectively. Here, these focal arrays are realized by multiplying two phases, yielding 1×4 focal arrays along the x- and y-axis, respectively. Thus, $N=4$, $z_i=0$, and $\delta_i=0$. For 1×4 foci along the x-axis, $y_i=0$, with the foci located at $x_1=-3$, $x_2=3$, $x_3=-9$ and $x_4=9$. For 1×4 foci along the y-axis, $x_i=0$, with the foci located at $y_1=-3$, $y_2=3$, $y_3=-9$ and $y_4=9$. Three pupil filters for the 1×4 focal array along the x- and y-axis are obtained, with (i) $s_1=1.05, s_2=0.7, s_3=0.92, s_4=-0.28$; (ii) $s_1=s_2=-0.885$, $s_3=s_4=1$; and (iii) $s_1=-1.05, s_2=0.7, s_3=0.95, s_4=0.33$, which are denoted as $l_{x1} \sim l_{x3}$ for the x-axis and $l_{y1} \sim l_{y3}$ for the y-axis. Finally, the phases for (d-f) are (d) $l_{x1} \times l_{y1}$; (e) $l_{x2} \times l_{y2}$; and (f) $l_{x3} \times l_{y2}$, respectively.

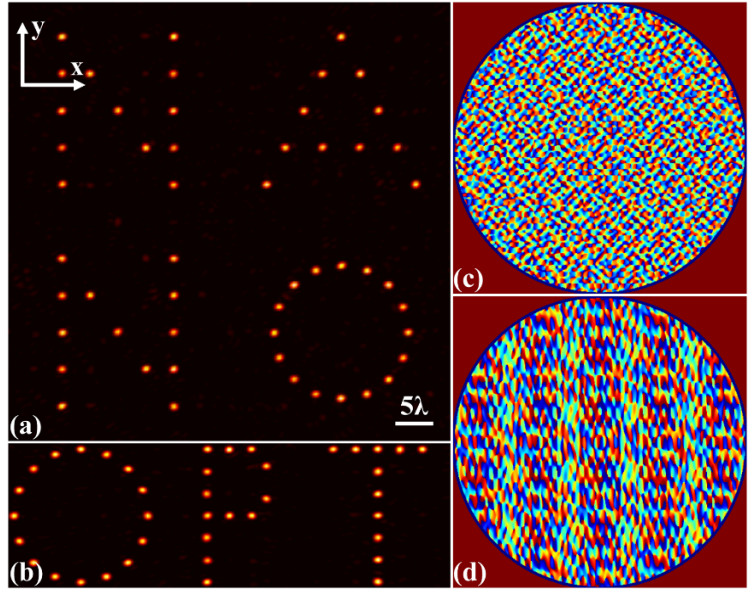

**Figure S11** "NANO" (a) and "OPT" (b) are generated in the focal plane using the phases in (c) and (d), respectively. All parameters can be found in the Appendix.

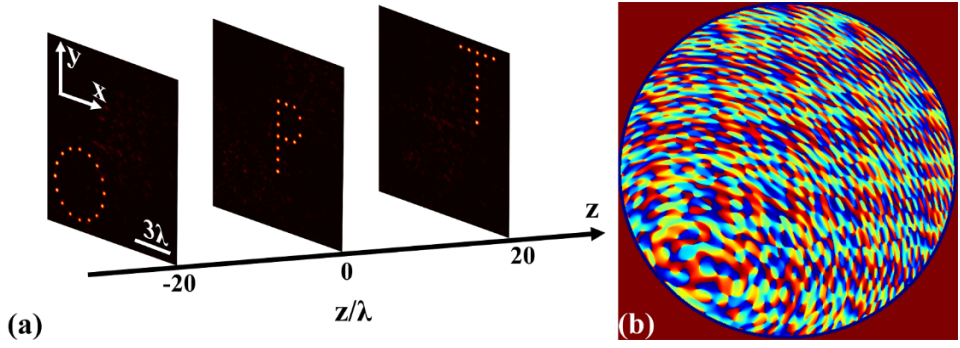

**Figure S12** "O", "P", and "T" are generated in the z=–20, z=0 and z=20 planes simultaneously using the phase in (b). The parameters can be found in the Appendix.

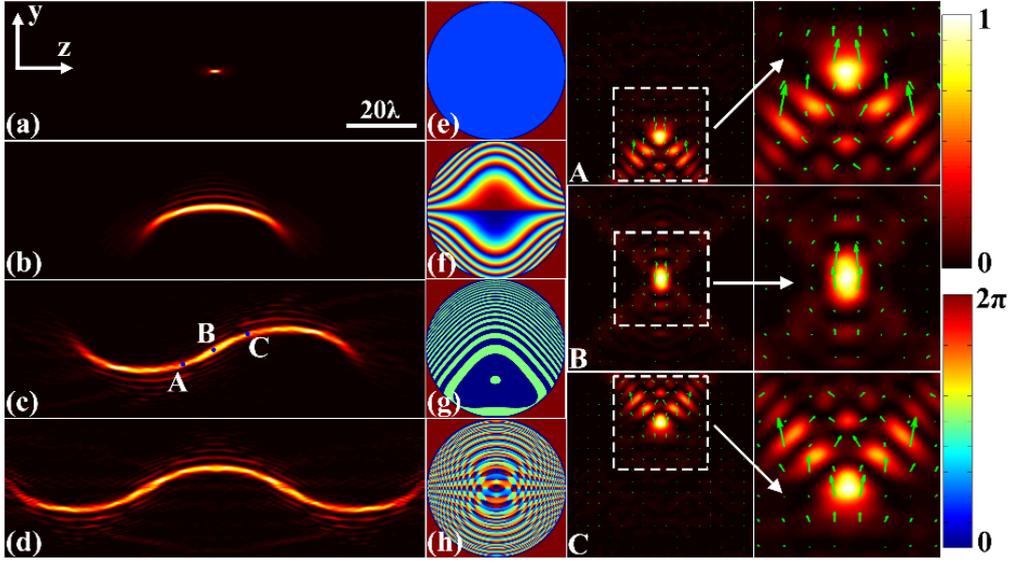

**Figure S13** UAD light beams with different numbers of energy oscillations under the condition of *NA*=0.8: (a) zero oscillations (2.34λ); (b) one oscillation (31.96λ); (c) two oscillations (70.56λ); (d) three oscillations (99.50λ); (e-h) are their corresponding phases, the parameters of which can be found in Table S1; (A-C) present the light intensities and energy fluxes (Green arrows) of points A, B, and C in the z=-10λ, 0, 10λ planes, respectively.

| N | $y_1$ | $y_2$ | $y_3$ | $z_1$ | $z_2$ | $z_3$ |
|---|---|---|---|---|---|---|
| 1 | 0 | | | 0 | | |
| 1 | 0 | | | 0 | | |
| 2 | -4.92 | 4.92 | | -19.505 | 19.505 | |
| 3 | -4.92 | 4.92 | -4.92 | -39.12 | 0 | 39.12 |
| $\eta_1$ | $\eta_2$ | $\eta_3$ | $\delta_1$ | $\delta_2$ | $\delta_3$ | Parameter |
| 0 | | | | | | $NA = 0.8$ |
| 5 | | | | | | $\phi_0 = 0.75\pi$ |
| -5 | 5 | | $-0.95\pi$ | $0.95\pi$ | | $x_i = 0$ |
| -5 | 5 | -5 | $0.26\pi$ | 0 | $-0.26\pi$ | $s_i = 1$ |

Table S1: UAD Light Beams of Figure S13